\documentclass{svproc}

\usepackage{siunitx}
\usepackage{amsmath,bm}
\usepackage{amsfonts}
\usepackage{mathtools}
\usepackage{multirow}

\usepackage{url}

\newcommand{\stiffAZ}{\mathbf{A}}
\newcommand{\massAZ}{\mathbf{B}}
\newcommand{\stiffAZtc}{\hat{\stiffAZ}}
\newcommand{\massAZtc}{\hat{\massAZ}}
\newcommand{\stiffAZred}{\tilde{\stiffAZ}}
\newcommand{\massAZred}{\tilde{\massAZ}}
\newcommand{\lambdatcAZ}{\hat{\lambda}}
\newcommand{\lambdaredAZ}{\tilde{\lambda}}
\newcommand{\zetasAZ}{\mathbf{Z}}

\newcommand{\evecAZ}{\mathbf{v}}
\newcommand{\evecAZtc}{\hat{\evecAZ}}
\newcommand{\evecAZred}{\tilde{\evecAZ}}

\newcommand{\NpodAZ}{N_{\mathrm{POD}}}
\newcommand{\NinitAZ}{N_{\mathrm{init}}}
\newcommand{\NfeAZ}{N_{\mathrm{IGA}}}
\newcommand{\NtcAZ}{|\mathrm{C}|}
\newcommand{\NredAZ}{N_{\mathrm{red}}}
\newcommand{\NtrainAZ}{N_{\mathrm{train}}}
\newcommand{\snapshotAZ}{\mathbf{Y}}
\newcommand{\HBinvAZ}{\left(\mathbf{H}(t)\massAZ(t)^{-1}\right)}
\newcommand{\NmaxAZ}{N_{\mathrm{max}}}
\newcommand{\nEVAZ}{K}

\newcommand{\changeAZ}[1]{\textcolor{black}{#1}}
\newcommand{\changeAZtwo}[1]{\textcolor{black}{#1}}

\usepackage{subcaption}

\begin{document}
\mainmatter              
\title{A Mixed Tree-Cotree Gauge for the Reduced Basis Approximation of Maxwell's Eigenvalue Problem}
\titlerunning{A Mixed Tree-Cotree Gauge}  
%
\author{Anna Ziegler \and Sebastian Schöps
\authorrunning{Anna Ziegler, Sebastian Schöps} 
%
%
\institute{Computational Electromagnetics Group, Schloßgartenstr. 8, 64289 Darmstadt, Germany,\\
\email{anna.ziegler@tu-darmstadt.de, sebastian.schoeps@tu-darmstadt.de},\\
\texttt{https://www.cem.tu-darmstadt.de/}}}

\maketitle              

\begin{abstract}
Model order reduction methods are a powerful tool to drastically reduce the computational effort of problems which need to be evaluated repeatedly, i.e., when computing the same system for various parameter values.
When applying a reduced basis approximation algorithm to the Maxwell eigenvalue problem, we encounter spurious solutions in the reduced system which hence need to be removed during the basis construction. 
In this paper, we discuss two tree-cotree gauge-based methods for the removal of the spurious eigenmodes.
\end{abstract}
\section{Introduction}
Model order reduction methods aim at condensing large models to smaller ones, which can be evaluated in an acceptable amount of time and using limited memory, but still obtain a
reliable outcome~\cite{Schilders_2008ab_AZ}.
In previous works~\cite{Ziegler_2024aa_AZ}, we have introduced a reduced basis approximation based on \cite{Horger_2017aa_AZ} for the Maxwell eigenvalue problem on parametrized domains. 
We have proposed three methods to remove spurious solutions from the basis.
One idea is based on the tree-cotree gauge as presented by Manges et al.~\cite{Manges_1995aa_AZ}.
Although this reliably removes the spurious modes and does not compromise a high accuracy of the reduced basis, it comes with computationally inconvenient properties.
Therefore, in this work, we want to adapt this classical tree-cotree gauge in a way, that still benefits from the accuracy and reliability, but allows for faster computation and less memory consumption. 

In the following, we will first state the continuous electromagnetic eigenvalue problem and its discretized counterpart.
In Sec.~\ref{sec:ziegler_rb}, we briefly recall the used model order reduction method and discuss the occurrence of the spurious eigenmodes in the reduced basis.
Then, we will discuss the classical tree-cotree gauge and our adapted approach in detail.
Finally, we will address some computational considerations and  present numerical results in Sec.~\ref{sec:ziegler_numerics}, and conclude our work in Sec.~\ref{sec:ziegler_conclusion}.

\subsection{Problem Statement}
We consider the Maxwell eigenvalue problem in an electromagnetic cavity, i.e., on a bounded, simply connected domain $\Omega\in\mathbb{R}^3$ in vacuum with Lipschitz boundary $\partial \Omega$.
We set $\mathbf{E}\times\mathbf{n}=0$ on $\partial \Omega$, where by $\mathbf E$ we denote the electric field strength, by $\mathbf{n}$ the outwards pointing normal vector and use the source-free time-harmonic formulation 
\begin{equation}
\nabla\times\left(\nabla\times\mathbf{E}\right)={\omega^{2}}/{c_{0}^2}\mathbf{E}  \quad\text{in }\Omega
\label{eq:ziegler_maxwellProblem}
\end{equation}
with $c_0$ the speed of light in vacuum and $\omega$ the angular resonance frequency of the cavity.
For the discretization of problem~\eqref{eq:ziegler_maxwellProblem}, we use isogeometric analysis (IGA), where we use B-splines and NURBS as basis functions for analysis and geometry~\cite{Vazquez_2010aa_AZ}.
In our work, we parameterize the domain $\Omega(t)$ and its boundary, therefore we introduce the deformation parameter~$t \in T = [0,1]$.
Thereby, we obtain the discrete, generalized eigenvalue problem
\begin{equation} 
\stiffAZ(t)\evecAZ(t) = \lambda(t) \massAZ(t) \evecAZ(t)
\label{eq:ziegler_eigproblem}
\end{equation} 
of dimension~$\NfeAZ$ with parameter-dependent eigenvector~$\evecAZ(t)$ and eigenvalue~$\lambda(t)$.

\section{Reduced Basis Approximation \changeAZ{and Spurious Modes}}
\label{sec:ziegler_rb}
To build a reduced model approximating the first $K$ eigenvalues of interest, we follow the reduced basis approach from Horger et al.~\cite{Horger_2017aa_AZ}.
In their work, the authors initialize the reduced basis with proper orthogonal decomposition (POD)~\cite{Kahlbacher_2007aa_short_AZ} and enhance the basis following a greedy strategy by adding the eigenvector, which is approximated with the least accuracy by the current reduced basis according to an error estimator.
The adaption of the algorithm for the Maxwell eigenvalue problem and parameter-dependent domains is described in more detail in~\cite{Ziegler_2024aa_AZ}.

The algorithm uses a set of eigenvectors of the high-fidelity system \eqref{eq:ziegler_eigproblem} as inputs. From those, it computes the reduced basis matrix~$\zetasAZ$, by which we obtain the reduced stiffness and mass matrices
\begin{equation}\label{eq:ziegler_reduceMatrices}
\stiffAZred(t) \coloneqq \zetasAZ^\top \stiffAZ(t)\zetasAZ,
\quad
\massAZred(t) \coloneqq \zetasAZ^\top \massAZ(t)\zetasAZ.
\end{equation}
After solving the reduced eigenvalue problem on $\stiffAZred(t)$ and $\massAZred(t)$, we can use the matrix~$\zetasAZ$ to upscale the reduced eigenvector $\evecAZred(t)$ to the original space by the matrix-vector product 
$
\evecAZ(t) \approx \zetasAZ \evecAZred(t).    
$
Thereby, we can obtain the eigenvalues and eigenvectors of the original system with the desired accuracy but with much less effort. 
We will go into further details on the basis computation when we present our novel mixed gauged approach in Sec.~\ref{sec:ziegler_mixedGauge}.

When applying the reduced basis approximation to the Maxwell eigenvalue problem, one encounters spurious eigenmodes in the reduced basis.
These appear as a non-trivial nullspace of the curl-curl-operator, i.e., as gradient fields.
\changeAZtwo{This was also observed in \cite{Rubia_2022aa_AZ}.}
We can use various methods to remove these gradient fields, as shown in~\cite{Ziegler_2024aa_AZ}. 
One option is the orthogonalization of the basis vectors with respect to the gradient space.
However, it showed that these methods are susceptible to numerical errors. 
From the investigated approaches, we found the tree-cotree gauge-based method following Manges et al.~\cite{Manges_1995aa_AZ} to be most successful and reliable. 
Nevertheless, also this approach does not come without drawbacks. 
Before we present our adjustments to the algorithm in order to deal with these limitations, we first recall and discuss the classical tree-cotree gauge.

\subsection{Tree-Cotree Gauge}\label{sec:ziegler_treecotree}
In order to use the tree-cotree gauge by Manges et al., we find a spanning tree and its cotree on the edges of the IGA control mesh of the domain, which are associated with the degrees of freedom (DoFs), when using Nédélec-type basis functions.
We denote the set of indices belonging to the tree and cotree edges with~$T$ and~$C$, respectively, and define the matrix
    $\mathbf{H} \coloneqq [\stiffAZ_{CC}, \stiffAZ_{CT}]$.
We then formulate the problem on the cotree DoFs by defining the matrices 
\begin{equation}\label{eq:ziegler_cotreeMatrices}
\stiffAZtc = (\mathbf{H}\massAZ^{-1})\stiffAZ(\mathbf{H}\massAZ^{-1})^\top, \quad \text{and} \quad 
\massAZtc = (\mathbf{H}\massAZ^{-1})\massAZ(\mathbf{H}\massAZ^{-1})^\top
\end{equation}
and perform a transformation of variables  via 
$
\evecAZ = \massAZ^{-1}\mathbf{H}^\top\evecAZtc,
$
such that we obtain the new cotree system 
\begin{equation}
    \stiffAZtc\evecAZtc = \lambdatcAZ\massAZtc\evecAZtc
    \label{eq:ziegler_cotreeSystem}
\end{equation}
of dimension~$\NtcAZ < \NfeAZ$.
The solution of \eqref{eq:ziegler_cotreeSystem} excludes gradients and thus could be used directly to build the reduced basis.
Unfortunately, system~\eqref{eq:ziegler_cotreeSystem} is numerically inconvenient, as it is given by dense matrices with elevated condition numbers \cite{Munteanu_2002aa_AZ}.
Due to this, we experience a rather high memory consumption as well as an increased computational effort for solving the eigenvalue problem.
Therefore, we propose an adaption to circumvent the explicit assembly and solution of the cotree system matrices.

\subsection{Mixed Gauged Variant}\label{sec:ziegler_mixedGauge}
\begin{figure}[!htb]
    \centering
    \includegraphics[scale = 0.99]{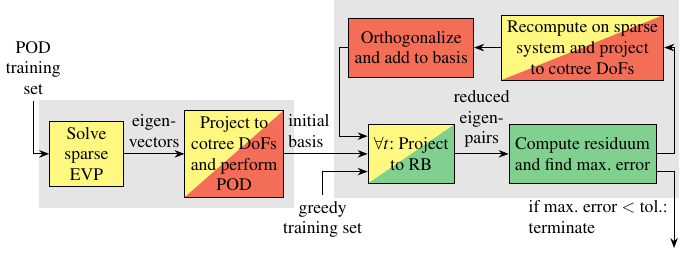}
    \caption{Flowchart of the mixed gauged algorithm. The colors indicate the computational space.  
    \changeAZ{Yellow corresponds to the sparse IGA space,
    red to dense matrices of the cotree space} and green to the reduced space. The two phases are indicated with the gray background boxes.}
    \label{fig:ziegler_flowchart}
\end{figure}

The considered reduced basis algorithm consists of two phases, namely an initialization phase and a second phase, where the reduced basis is enriched in a greedy manner. 
Since these phases follow different procedures, we will discuss them separately.
The algorithm is illustrated in the flowchart in Fig.~\ref{fig:ziegler_flowchart}.

\subsubsection{Initialization of the reduced basis}
Our idea for advancing the first phase is based on the objective of avoiding solving the eigenvalue problem on the cotree system matrices.
We start this phase by defining an initial training set $\Xi_{\mathrm{train}}^{\mathrm{POD}}
=\{t_1,\ldots,t_{\NpodAZ}\}$ of size $\NpodAZ$.
For this set, we assemble the corresponding system matrices~$\stiffAZ(t)$ and~$\massAZ(t)$.
Instead of directly transferring these to the cotree system matrices via~\eqref{eq:ziegler_cotreeMatrices}, we apply the variable transformation leading to \eqref{eq:ziegler_cotreeSystem} 
in a reverse way, i.e., we solve the sparse eigenvalue problem~\eqref{eq:ziegler_eigproblem} and collect a set of solution eigenvectors.
On this selection, we then apply the transformation
\begin{equation}
    \evecAZtc(t) = \mathbf{H}(t)^{-\top} \massAZ(t) \evecAZ(t)
\end{equation}
to condense the eigenvectors to the cotree DoFs and collect these vectors~$\evecAZtc(t)$ in a snapshot matrix~$\snapshotAZ$. 
\changeAZ{Thereby}, we avoid solving the cotree system and solve only the original system, which has more DoFs but is sparse, and faster to solve. 
Nevertheless, we have removed the non-trivial nullspace where a spurious gradient field could occur. 
On this snapshot matrix, we apply the POD which decomposes the snapshots according to a singular value decomposition.
By ordering the obtained components according to the size of the singular values, we can select the $\NinitAZ$ most dominant modes of the snapshots which capture the most significant features, e.g. \cite{Kahlbacher_2007aa_short_AZ}, and add these into the initial matrix $\mathbf{Z} \coloneqq ({\mathbf{z}_1},\ldots, \mathbf{z}_{\NinitAZ})$ where each $\mathbf{z}_i \in \mathbb{R}^{\NtcAZ}$ has the dimension of the cotree edges DoFs.
Thereby, we obtain an initial basis that we improve in the second phase of the algorithm. 
\subsubsection{Greedy enrichment of the reduced basis}
After having computed an initial basis, it is enhanced by iteratively adding eigenvectors to the basis.
These vectors are chosen following an error estimator to identify that eigenvector, approximated the least accurately by the current basis.
The error estimator as proposed by Horger et al.~\cite{Horger_2017aa_AZ} takes the eigenvalue computed on the reduced basis, its distance to neighboring eigenvalues, and the residuum with respect to the high-fidelity solution into account.
Since the residuum needs to be evaluated for each value in the new training set~$\Xi_{\mathrm{train}}$ of $\NtrainAZ$ parameters for the greedy phase of the algorithm, an efficient procedure is required.
Instead of evaluating the eigenpairs on the high-fidelity system, we project the original system matrices to the reduced space to solve the reduced eigenvalue problem. To this end, we combine~\eqref{eq:ziegler_cotreeMatrices} and~\eqref{eq:ziegler_reduceMatrices} to
\begin{equation}\label{eq:ziegler_reducedMatricesMixed}
\begin{aligned}
    \stiffAZred(t) &= \zetasAZ^\top \HBinvAZ \stiffAZ(t) \HBinvAZ^\top \zetasAZ,\\
    \massAZred(t) &= \zetasAZ^\top \HBinvAZ \massAZ(t) \HBinvAZ^\top \zetasAZ.
\end{aligned}
\end{equation}
By a clever implementation, the projection to the reduced space can be performed efficiently, since $\mathbf{Z} \in \mathbb{R}^{\NtcAZ \times \NredAZ}$ where $\NredAZ \ll \NtcAZ$.
Due to the small size of the system, we can then solve the reduced eigenvalue problem with $\stiffAZred(t)$ and $\massAZred(t)$ rapidly for $\lambdaredAZ$ and $\evecAZred$ and compute the residuum with respect to the high-fidelity solution via
\begin{equation}\label{eq:ziegler_residuum}
\mathbf{r}_{i}(t) = \stiffAZ(t)  \left(\massAZ^{-1}(t) \mathbf{H}(t)\changeAZtwo{^\top}\right)  \zetasAZ \evecAZred_i(t)  - \lambdaredAZ\changeAZtwo{_i}(t)  \massAZ(t)  \left(\massAZ^{-1}(t) \mathbf{H}(t)\changeAZtwo{^\top} \right) \zetasAZ \evecAZred_i(t),
\end{equation}
which does not require the solution of the high-fidelity problem.
Using the residuum, we evaluate the error estimator and determine the maximum error and the corresponding eigenpair.
For that eigenpair, we recompute the solution on the high-fidelity system, project it to the cotree space, orthogonalize it towards the current reduced basis and then add it to the basis.
We terminate the algorithm once the maximum error is smaller than a user-defined tolerance or when the maximum size $\NmaxAZ$ of the basis is reached.
\section{Computational Considerations and Numerical Results}\label{sec:ziegler_numerics}
All implementations are carried out in MATLAB\textsuperscript{\textregistered} R2022a using  GeoPDEs~\cite {Vazquez_2016aa_AZ} for the discretization. 
The implementation uses $C^1$-continuous second-order spline edge basis functions and the tree is constructed as proposed in \cite{Kapidani_2022aa}.
A workstation with an Intel\textsuperscript{\textregistered} CPU i7-3820 3.6-GHz processor and 24 GB RAM is used.

\textbf{Memory} For all parameters in the POD training set, we solve the sparse high-fidelity system and project the eigenvector to the cotree space.
To this end, we use the MATLAB\textsuperscript{\textregistered} backslash-operator or equivalently the \texttt{mldivide} function on the first~$K$ sparse eigenvectors $\evecAZ$, i.e., \texttt{\^{v} = H$^\top$ $\backslash$ (B * v)}.
This avoids the costlier computation of $\mathbf{H} \massAZ^{-1}$ as \texttt{H / B} (or equivalently using \texttt{mrdivide}) to obtain the cotree matrices and solving the eigenvalue problem on these, as well as keeping the dense matrices in the memory. 

In the greedy phase of the algorithm, once we have an initial basis $\mathbf{Z}$, we move the multiplication with this ahead of further computations to immediately reduce their dimensions.  
For this purpose, we define an additional projection matrix
\begin{equation} \label{eq:ziegler_Ztc}
    \hat{\mathbf{Z}}^\top \coloneqq \mathbf{Z}^\top \mathbf{H} \massAZ^{-1}
\end{equation}
and obtain the reduced matrices via
\begin{equation}\label{eq:ziegler_reduceMatricesCotree}
\stiffAZred(t) = \hat{\zetasAZ}^\top \stiffAZ(t)\hat{\zetasAZ},
\quad
\massAZred(t) = \hat{\zetasAZ}^\top \massAZ(t)\hat{\zetasAZ},
\end{equation}
by computing \texttt{\^{Z}$^\top$  = (Z$^\top$ * H) / B}.
Since we need to repeat this computation for each instance of the training set and for each iteration of the greedy algorithm, we speed up the calculation of~\eqref{eq:ziegler_Ztc} using a sparse Cholesky decomposition via the Matlab function \texttt{decomposition()}.
Then, we solve the reduced systems.
Equivalently, for the computation of the residuum~\eqref{eq:ziegler_residuum}, we ensure dimension-beneficial multiplication ordering and use the matrix decomposition.
These adaptations drastically reduce the memory consumption since the dense cotree matrices are not explicitly computed and stored.

\begin{figure}
    \centering
    \begin{subfigure}[c]{0.33\textwidth}
        \centering
        \includegraphics[width = \textwidth, angle = 0]{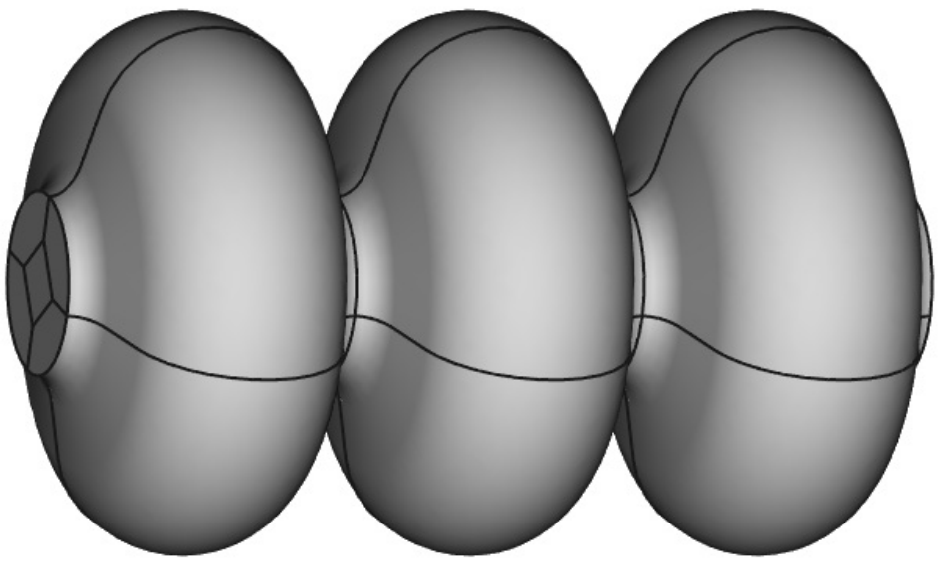}~%
    \end{subfigure}~%
     \begin{subfigure}[c]{0.33\textwidth}
        \centering
        \includegraphics[width = 0.38\textwidth, angle = 0]{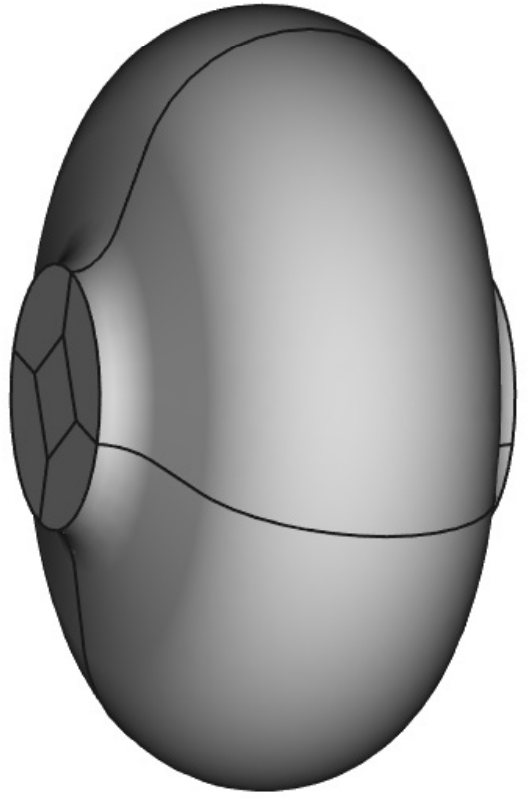}~%
    \end{subfigure}~%
     \begin{subfigure}[c]{0.25\textwidth}
        \centering
        \includegraphics[width = 0.8\textwidth, angle = 0]{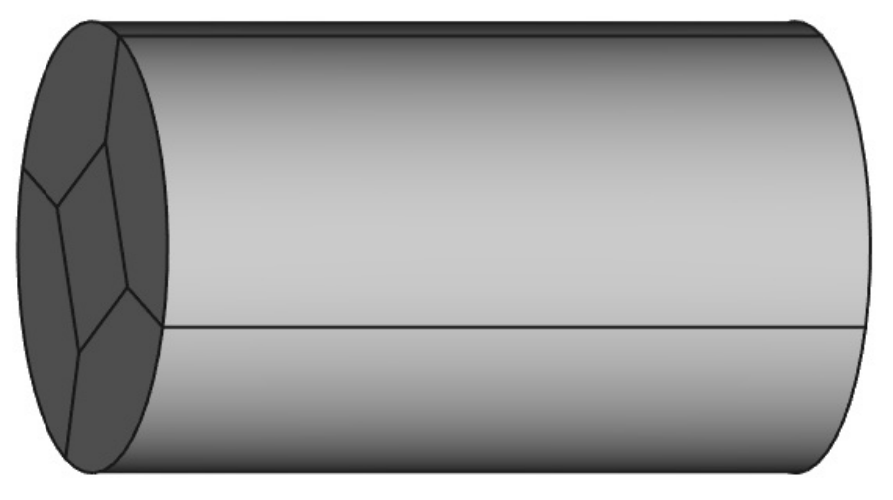}
    \end{subfigure}
    \caption{\changeAZ{3-cell and 1-cell TESLA cavity at $t = 0$ (left, middle) morphed to the Pillbox cavity at $t=1$ (right). For $t \in (0,1)$, the system matrices are obtained via matrix interpolation and do not correspond to physical geometries.}}
\end{figure}

\begin{figure}[htb]
\centering
\includegraphics[]{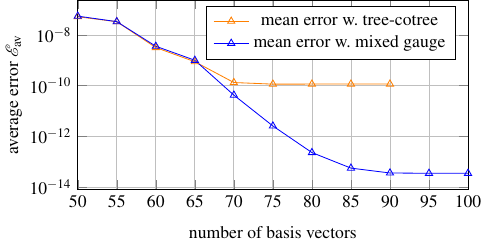}
\caption{Average error of the first five approximated eigenvalues in the one-cell TESLA cavity with 50 initial basis vectors.}
\label{fig:ziegler_errors}
\end{figure}

\textbf{Accuracy} Furthermore, solving the eigenvalue problem on the sparse system allows for higher accuracy.
We demonstrate this with an example, where 
for the parameterized domain, we choose the elliptic TESLA cavity at $t=0$ and the cylindrical pillbox cavity at $t=1$ and model intermediate structures via convex combinations of the system matrices along~$t$.  
In order to avoid solving on the full system of $2148$ DoFs, we build the reduced basis by training it on $50$ and $100$ values of $t$ in the POD and the greedy procedure, respectively. 
We then evaluate the error on $200$ randomly chosen systems and compute the average relative error $\mathcal{E}_{i,\mathrm{av}}$ of the eigenvalues~$i = 1,\ldots, 5$ w.r.t. the high-fidelity solution, c.f. Fig.~\ref{fig:ziegler_errors}.
While the basis using the classical tree-cotree gauge then converges to the error of the tree-cotree gauged high-fidelity system of approx.  $1.16 \cdot 10^{-10}$, the reduced basis using the mixed gauge further improves in accuracy until a size of approximately $90$ basis vectors where it converges to an error of $3.50 \cdot 10^{-14}$ of the IGA discretization. 
 
\textbf{Runtime} In the next step, we investigate the runtime efficiency for the construction of the basis as well as for the speedup in the evaluation of eigenvalue problems. 
For the comparison, we again use the example of the one-cell TESLA cavity with $2148$ DoFs, which is deformed to the pillbox cavity. 
Along the deformation, we track the eigenmodes with the algorithm from \cite{Ziegler_2023aa_AZ} and ensure a high consistency when matching the modes along the iterations with a minimum correlation of the modes of $90\%$, possibly adaptively requiring rather small tracking steps.
As examples with more DoFs, we consider a finer discretization of the one-cell cavity with $5248$ DoFs and also the  equivalent deformation, starting from the three-cell TESLA cavity.
In all settings, we train on $20$ and $50$ instances in the POD and the greedy training sets, respectively.
For approximating $\nEVAZ = 5$ eigenvalues of interest of the one-cell TESLA cavity, we choose a basis of size $\NmaxAZ = 75$ for the coarser discretization. 
For the finer mesh, we use a reduction to $\NmaxAZ = 80$ DoFs.
For the approximation of $\nEVAZ = 15$ in the three-cell cavity, we reduce the original $6612$ DoFs to $210$.
We initialize the bases with $\NinitAZ = 65$ vectors chosen from the POD for the one-cell geometries and $160$ vectors for the three-cell design.
We remark that in case of lower accuracy requirements, also smaller basis sizes might be sufficient. 

Let us point out that our code is not runtime-optimized, but has prototype character. Our time measurements should therefore be seen more as qualitative indicators.
In Tab.~\ref{tab:ziegler_runtimeResults}, we first compare the performance of the classical tree-cotree gauged tracking with the mixed gauged tracking on the coarser model of the one-cell cavity.
With the mixed gauge, the construction of the basis and tracking on the reduced problem (including the matrix assemblies) takes approx. \SI{26}{\s} and therefore is more efficient than the tracking on the high-fidelity problem, which takes \SI{48}{\s}.
This is in contrast to the variant using the classical gauge, where the entire procedure takes \SI{117}{\s} due to the elevated time for solving the high-fidelity eigenvalue problem on the dense matrices and the effort for assembling the cotree matrices.
\changeAZ{This trend continues as the systems become larger, making our mixed gauge clearly the preferred variant.} 
In the examples with more DoFs, we observe the \changeAZ{mixed gauged} reduced basis to become increasingly profitable. 
While for the coarser one-cell model, the speedup for the full procedure (with the mixed gauge) is $1.9$, in the finer discretization it increases to $2.9$ and to $24.8$ for the three-cell model.

\begin{table}[htb]
    \caption{Comparison of the reduced bases for different examples, obtained with the classical tree-cotree gauge and the mixed gauge.}
    \label{tab:ziegler_runtimeResults}
    \centering
    \begin{tabular}{l||c|c||c|c||c|c}
    \multirow{2}{*}{}     &  \multicolumn{2}{c||}{1 cell} & \multicolumn{2}{c||}{1 cell}  & \multicolumn{2}{c}{3 cells}  \\ 
       & Classic & Mixed  & \changeAZ{Classic}  & Mixed  & \changeAZ{Classic}  & Mixed \\ \hline
    \changeAZ{DoFs of original system} & \multicolumn{2}{c||}{\changeAZ{$2148$}} & \multicolumn{2}{c||}{\changeAZ{$5248$}}  & \multicolumn{2}{c}{\changeAZ{$6612$}}  \\ 
    \changeAZ{DoFs of reduced system} & \multicolumn{2}{c||}{\changeAZ{$75$}} & \multicolumn{2}{c||}{\changeAZ{$80$}}  & \multicolumn{2}{c}{\changeAZ{$210$}}  \\ 
    \changeAZ{No. of approx. EVs $K$} & \multicolumn{2}{c||}{\changeAZ{$5$}} & \multicolumn{2}{c||}{\changeAZ{$5$}}  & \multicolumn{2}{c}{\changeAZ{$15$}}  \\
    \changeAZ{Initial basis $\NinitAZ$} & \multicolumn{2}{c||}{\changeAZ{$65$}} & \multicolumn{2}{c||}{\changeAZ{$65$}}  & \multicolumn{2}{c}{\changeAZ{$160$}}  \\
    Projection to Cotree DoFs & \SI{36.1}{\s}  & \SI{6.5}{\s} & \changeAZ{\SI{7}{\min}  \SI{31}{\s}} &\SI{33.4}{\s} & \changeAZ{\SI{14}{\min}  \SI{41}{\s}} & \SI{2}{\min}  \SI{34}{\s} \\
    POD & \SI{21.6}{\s} & \SI{7.6}{\s} & \changeAZ{\SI{2}{\min} \SI{03}{\s}} &\SI{34.1}{\s}& \changeAZ{\SI{3}{\min} \SI{15}{\s}} & \SI{33.2}{\s}  \\
    Greedy & \SI{11.6}{\s} & \SI{10.5}{\s} & \changeAZ{\SI{2}{\min} \SI{52}{\s}} &  \SI{50.8}{\s} & \changeAZ{\SI{18}{\min} \SI{9}{\s}} &\SI{5}{\min} \SI{4}{\s}  \\  
    Tracking (RB) & \SI{47.7}{\s} &  \SI{1.3}{\s}  & \changeAZ{\SI{35}{\min}  } &\SI{5.0}{\s} & \changeAZ{\SI{9}{\hour}} &\SI{2}{\min} \SI{9}{\s}\\ \hline
    EVP (full, cotree/sparse) & \SI{0.5294}{\s} & \SI{0.1132}{\s} & \changeAZ{\SI{1.49}{\s}} &\SI{0.3678}{\s} & \changeAZ{\SI{4.50}{\s}} & \SI{0.6321}{\s} \\ 
    EVP (RB) & \SI{0.0008}{\s} & \SI{0.0008}{\s} & \changeAZ{\SI{0.0009}{\s}} &\SI{0.0008}{\s} & \changeAZ{\SI{0.0056}{\s}} & \SI{0.0053}{\s}\\ 
    Tracking (full, sparse) & \multicolumn{2}{c||}{\SI{48.4}{\s}} & \multicolumn{2}{c||}{\SI{5}{\min}  \SI{56}{\s}}  & \multicolumn{2}{c}{\SI{4}{\hour} \SI{16}{\min}} \\ 
   
    \end{tabular}
\end{table}

\section{Conclusion}\label{sec:ziegler_conclusion}
We proposed an improvement of the tree-cotree gauge-based removal of spurious modes for the reduced basis approximation of Maxwell's eigenvalue problem.
When tracking eigenvalues this method outperforms the classical tree-cotree approach in the construction of the basis, as well as in the application phase.
The construction and application of the reduced basis allowed a speedup compared to the high-fidelity tracking procedure up to approx. 25 times. 
In future work, the method can also be integrated into an `online-greedy' algorithm by immediately starting the tracking phase after initialization with POD.
Then, further vectors are added to the basis according to the error estimator in each tracking step using the greedy principle only as required.

\paragraph{Acknowledgement}
This work is supported by the Graduate School CE within Computational Engineering at Technische Universität Darmstadt.

\bibliographystyle{spmpsci}

\end{document}